# Whispering-Gallery-Mode Resonators for Detection and Classification of Free-Flowing Nanoparticles and Cells through Photoacoustic Signatures

Jie Liao[1], Maxwell Adolphson[1], Hangyue Li[2], Dipayon Kumar Sikder[1], Chenyang Lu[1,2,3,4], Lan Yang[1,4]

[1]Department of Electrical and Systems Engineering, Washington University, St. Louis, MO 63130, USA
[2]Department of Computer Science and Engineering, Washington University, St. Louis, MO 63130, USA
[3]Department of Medicine, Washington University, St. Louis, MO 63130, USA
[4]AI for Health Institute, Washington University, St. Louis, MO 63130, USA
E-mail: yang@seas.wustl.edu

## Abstract

Micro and nanoscale particles have played crucial roles across diverse fields, from biomedical imaging and environmental processes to early disease diagnosis, influencing numerous scientific research and industrial applications. Their unique characteristics demand accurate detection, characterization, and identification. However, conventional spectroscopy and microscopy commonly used to characterize and identify tiny objects often involve bulky equipment and intricate, time-consuming sample preparation. Over the past two decades, optical micro-sensors have emerged as a promising sensor technology with their high sensitivity and compact configuration. However, their broad applicability is constrained by the requirement of surface binding for selective sensing and the difficulty in differentiating between various sensing targets, which limits their application to detecting targets in their native state or complex biological samples. Developing label-free and immobilization-free sensing techniques that can directly detect target particles in complex solutions is crucial for overcoming the inherent limitations of current biosensors. In this study, we design and demonstrate an optofluidic, high throughput, ultra-sensitive optical microresonator sensor that can capture subtle acoustic signals, analogous to whispers, generated by tiny particles from the absorption of pulsed light energy, providing photoacoustic spectroscopy information for real-time, label-free detection and interrogation of particles and cells in their native solution environments across an extended sensing volume. Leveraging unique optical absorption of the targets, our technique can selectively detect and classify particles flowing through the sensor systems without the need for surface binding, even in a complex sample matrix, such as whole blood samples. We showcase the measurement of gold nanoparticles with diverse geometries and different species of red blood cells in the presence of other cellular elements and a wide variety of proteins. These particles are effectively identified and classified based on their photoacoustic fingerprint that captures particle shape, composition, molecule properties, and morphology features. This work opens up new avenues to achieve rapid, reliable, and high-throughput particle and cell identification in clinical and industrial applications, offering a valuable tool for understanding complex biological and environmental systems.

## Introduction

Particles in micro and nanoscale are abundant in nature and play critical roles in diverse biological and environmental processes. Beyond their natural presence, these tiny particles are also present in a wide range of scientific research and industrial applications. Due to their small size, they possess unique physical, chemical, and biological characteristics markedly distinct from their bulk counterparts. Therefore, detecting, characterizing, and identifying these particles accurately and efficiently is crucial. In the field of medicine, for instance, nanoparticles are indispensable in targeted drug delivery systems, therapies, and diagnostics. They can interact with biological systems at the molecular level, enhancing treatment precision and efficacy[1,2]. In photothermal therapy, nanoparticles are used as agents that absorb light and convert it into heat, specifically targeting and destroying cancer cells[3,4]. In material science, nanomaterials exhibiting properties such as increased strength, chemical reactivity, or electrical conductivity open up new avenues for innovation in electronics, optics, and energy storage[5,6]. Given that the properties of particles depend on their composition, size, and shape, characterizing these particles becomes vital for understanding and leveraging the structure-property relationship[7].

Spectroscopic techniques like Raman, photoacoustic (PA), and UV-Vis spectroscopy are widely used for characterizing particles and molecules. Raman spectroscopy, for instance, relies on the Raman shift, which results from photons interacting with molecular vibrations and other low-frequency modes in a sample[8]. The Raman spectrum can provide a direct molecular fingerprint that reflects the composition of materials. However, the detection efficiency in Raman spectroscopy can be low, posing a challenge, especially for nanoscale objects[9]. Raman signals from water may also obscure the signals from the samples in an aqueous solution. In recent years, various Artificial Intelligence (AI) techniques have been explored to improve the performance of conventional sensing[10] and spectroscopy techniques[11]. PA spectroscopy combines optical excitation and acoustic detection, using short-pulsed lasers to probe and detect ultrasound generated by photon absorption and subsequent thermo-elastic expansion of targets. Analyzing the frequency domain features of photoacoustic signals provides insights into the shape, size, orientation of microstructures, and acoustic scattering properties, as well as various biophysical properties of the samples, making it a powerful tool for detailed characterization[12–16]. However, currently, this technique is mainly limited to nanoparticles with strong absorption properties, which are used as contrast agents for PA imaging[17], due to the lack of highly sensitive sensors capable of detecting very weak acoustic waves generated by nanoparticles on the nano- or micro-scales.

On the other hand, optical microsensors have emerged as a promising sensing technology in recent years, demonstrating great potential in small particle sensing due to their high sensitivity, rapid response, versatility, and affordability. The fundamental mechanism of detection involves analyzing alterations in optical signals as particles interact with light confined in the sensor. This provides information on various aspects of the target, such as its existence, concentration, and size[18–21]. Enhancing the light-matter interaction enables higher sensitivity; strategies including resonator-based sensing[22], plasmonic enhancement[23], and microlasers[24–26] have been widely exploited, demonstrating the most demanding biosensing tasks such as the detection of single virus particles and other nanoparticles[27–35]. However, in most cases, targets need to be in close proximity to or in direct contact with the micro-sensor's surface for detection. For example, optical resonator sensors use evanescent fields to probe the surrounding media, and plasmonic-enhanced sensors rely on "hot spots", posing a challenge in sample collection. Away from the surface, free-flowing particles in the sample solution remain undetected (Fig. 1a (i)). This means that only a small fraction of the target particles in the sample can be captured and analyzed, thereby limiting their detection efficiency and capacity for high-throughput sensing. In addition, measurements based on light intensity (phase) and resonance shift provide little identifiable differences in the properties of different molecules or particles. To achieve selectivity (specificity) in label-free sensing, the target molecules need to bind to receptors that are functionalized on the limited sensing surface. The need for surface binding and surface functionalization can introduce additional complexities and challenges, including surface fouling (non-specific binding), mass transport limitations (slow random diffusion rate, limiting the speed, sensitivity, and throughput), and surface regeneration (bound particles must be removed to reuse).

Moreover, to date, it's challenging for most optical microsensors, which are vulnerable to contamination, to obtain accurate measurements or detect specific targets in complex samples such as whole blood. The presence of numerous interfering substances, such as metabolites, electrolytes, and other biomolecules, can interact with the optical mode via evanescent fields, leading to false-positive or false-negative results, and compromising the specificity and sensitivity of the detection system. To ensure accuracy, they often necessitate the implementation of intricate, multi-step sample purification protocols, including incubation and washing. Such procedural complexities pose significant challenges to their practical deployment in clinical and industrial environments. In addition, the complex nature of the sample matrix can cause signal attenuation, optical absorption and scattering, or background noise, further obscuring the target signal and making it difficult to obtain accurate quantitative measurements. Therefore, there is a strong need for label-free sensing technologies that can detect target particles directly in complex solutions, without the need for surface binding and purification. Such techniques would enable the detection and analysis of target particles in their native environment, potentially leading to higher throughput, simpler and faster detection, improved sensitivity, and greater overall efficiency in biosensing applications.

Here, we demonstrate an optofluidic, high throughput, and label-free optical microresonator sensor with an extended detection volume, capable of real-time detection and interrogation of particles and cells in their native solution environments. The system is based on a high-quality ($Q$) bubble-shaped whispering gallery mode (WGMs) microresonator, featuring a hollow core directly connected to a microfluidic channel that allows sample solutions to flow inside while being optically stimulated by a pulsed laser from the outside to generate acoustic waves through PA effects. The acoustic waves carrying the PA spectroscopic information of the samples can propagate through the solution and reach the WGM sensor, which could be positioned either close or far from the site where the acoustic waves are initially generated. Therefore, our approach enables fast and simple detection and measurement of nanoscale objects across the microfluidic channel (Fig. 1a (ii)). It demonstrates, for the first time, the detection of free-flowing particles beyond the reach of the evanescent field close to the surface of an optical sensor under conventional detection schemes. It is also worth noting that the resonator's hollow core with a thick-wall design is ideally suited for integration with microfluidics. In this design, high-$Q$ WGMs are fully protected within the thick wall, avoiding any direct overlap between the sample and the optical field of the resonant mode. This optical isolation from the sample ensures that both the quality factor and the signal-to-noise ratio (SNR) remain unaffected by the absorbing and scattering properties of the sample. Since the PA process produces ultrasound in the MHz frequency range, the detection is background-free and is not affected by the environment. The exceptional stability and background-free detection of our novel sensing platform not only ensure reliable and consistent results across multiple measurements but also enable direct target detection in complex sample matrices, such as whole blood, without the need for laborious sample purification. We leverage this novel sensing platform to facilitate high-throughput photoacoustic identification of particles and cells in liquid media using advanced machine learning approaches. This approach enables sensitive and specific detection of particles and cells within a small sample volume (~μL), making it an ideal solution for applications requiring rapid, accurate, and cost-effective analysis in complex biological matrices.

## Results

### Optofluidic PA WGM sensor
The working principle of our sensor system is illustrated in Fig. 1b, which includes a WGM microbubble resonator (MBR)[29,36–38] built into a capillary to collect acoustic waves. The optical field of WGMs in the 780 nm band is confined within the silica wall. Solution samples containing particles flow through the fluidic channel connected to the MBR. When a particle flowing freely in the fluid is exposed to a pulsed laser at 532 nm, the energy absorbed by the particle causes a rapid thermoelastic expansion, resulting in the generation of an acoustic wave. These acoustic waves then propagate through the solution within the capillary to reach the WGM resonator. When an acoustic wave interacts with a WGM field, it alters the effective optical path length, and consequently modulates the optical resonance. In this way, PA signals carrying information about the particles can be read out from the continuous wave laser coupled to the microresonator. The acoustic wave generation by sensing targets, e.g., nanoparticles, via the PA process, and the detection of these acoustic waves by WGMs are two crucial features of the sensing process.

The key advantages of this technique are label-free and immobilization-free detection, significantly enhanced detection volume, and direct target detection in complex biological matrices. Different from conventional optical WGM sensors, which detect refractive index changes near the sensing surface (see a comparison in Supplementary Fig. S1), our sensing system detects acoustic waves generated by particles in solution flowing freely inside a capillary structure, which significantly improves the sensing capability, throughput, and speed. Furthermore, unlike conventional methods that rely on surface binding via random diffusion, we can actively scan the pulse laser across the microfluid channel and search for target particles. As shown in Fig. 1b (i)-(iii), by scanning along the capillary, the particles far away from the MBR where the WGM resides can also be detected. The resulting PA signal exhibits a delay between the pulse excitation and its optical readout, due to the distance between the optical mode and the particles. We experimentally measured the PA signals from particles at various distances from the MBR and achieved an extended detection volume (Supplementary Fig. S3). In addition, by selecting the wavelength of the pulse laser that overlaps with the optical absorption of the particle of interest, the target particles can be selectively and effectively detected even in a complex fluidic medium. As shown in Fig. 1c, a whole blood sample is injected into the sensor. In such a complex medium, red blood cells can be selectively detected in the presence of other cellular elements as well as proteins, due to the unique absorption wavelength of hemoglobin in red blood cells.

The specificity and reliability are improved by making our sensor immune to dispersive (refractive index changes) and dissipative changes (loss) in the solution. We design the geometry of the microresonator such that the radial mode field is strongly confined within the thick wall of the MBR (Supplementary Fig. S2). With little penetration into the core with the solution sample, the optical WGMs are protected from either refractive index changes or potential absorption or scattering loss in the sample. Even if the core is filled with black dye solution which has strong absorption, no apparent resonance shift or $Q$-factor deterioration is observed (Supplementary Fig. S2). Additionally, as we are detecting acoustic signals at high frequencies, the low-frequency noise (such as temperature drift and flow instability) is eliminated. In this way, the noise and false signal can be greatly mitigated and consequently improve reliability and robustness.

**Nanoparticle sensing**
To demonstrate the capability to detect and identify nanoparticles, four different geometries of gold nanoparticles (AuNPs) were tested in this study: nanospheres, nanorods, nanocubes, and nanoshells. Each solution was injected into the sensor via microfluidic channels. Fig. 2 shows typical photoacoustic spectra for each of the four geometries and their corresponding frequency domain signals. Accompanying each time and frequency domain plot are scanning transmission electron microscope (STEM) images of each geometry. Care was taken to excite each type of AuNPs at the same position in order to standardize the collection of each signal. The PA signal variations among the four geometries of AuNPs are notably distinct. This indicates that different shapes of nanoparticles, even when composed of the same material, produce unique PA signals. This spectral information in the PA signal further provides a measure of the shape property of the particles. Once we obtain a library of PA fingerprints for various nanoparticles, a spectral matching algorithm can be implemented to find the closest match between the unknown spectrum and the library spectra. The reference spectrum with the highest similarity score is considered the most likely match for the unknown particle to achieve particle classification.

**Detection of cells and identification from different species**
Next, we explored the feasibility of biological particle detection and identification using photoacoustic fingerprinting. Here we tested five different species of washed red blood cells—pig, sheep, turkey, goat, and llama. These cell suspension samples were diluted to 1%, by volume in a phosphate-buffered saline (PBS) solution, without any purification, labeling, or incubation process. They were then injected into the sensing system, and flown through the core of the capillary connected to the MBR. A pulse energy of ~40 nJ was used in the experiments to avoid photodamage. The temporal and frequency domain characterizations were obtained simultaneously, showing an SNR exceeding 30 dB. Fig. 3a-e shows the time-domain and frequency-domain PA signals of the five different sensing types of red blood cells. As shown in Fig. 3f, repeatability can be found in 10 measurements of red blood cells from the same species. In contrast with the results of AuNPs, the photoacoustic spectra of the 5 distinct cells look similar, with a major spectral peak near 4 MHz preserved across the species. Although there are some slight differences between different species, these differences may be not apparent to identify, and it could be challenging to distinguish the different species of cells directly. To tackle these challenges, we employed machine learning to analyze the differences at every single frequency component, which helped us identify them with high accuracy.

**Machine learning for cell classification from PA spectra**

To analyze the unique PA signals and extract the features associated with different species of red blood cells, we carried out comprehensive measurements, compiling a substantial dataset of PA signals for each species. Given the notable similarities in the PA signals and spectra across various species, it is crucial to identify the subtle characteristics of each type of blood cell. As a powerful tool to extract meaningful information from complex datasets, machine learning techniques were applied to learn from the PA signals by automatically identifying and extracting relevant features. The extracted features are more effective in distinguishing different types of red blood cells than the raw features.

To train the machine-learning-based feature extractor, we split the data into a training set (80%) and a testing dataset (20%). As shown in Fig. 3g, we first converted the input features from the time domain to the frequency domain using the Fast Fourier Transform (FFT). A one-dimensional convolutional neural network (CNN) is used to extract the spatial dependencies in the frequency domain of PA signals. The classification process begins by feeding the PA signals into the CNN, where the convolutional layer uses filters to extract features such as signal patterns. A pooling layer then reduces the data dimensionality, maintaining essential information as labeled in dashed red lines. The subsequent fully connected layer integrates these features, forming a comprehensive understanding to make predictions. To ensure robust predictions and mitigate the noise in signals, we further introduce prototype learning to enhance the similarity of the features extracted from the same cell species while reducing the chance of misclassification. The prediction depends on the sample's distance to the prototype embeddings in the latent space of each cell category[39]. The model calculates the shortest distance from the learned feature to the prototype embeddings and classifies the sample as the cell category of its closest prototype embedding (see Methods and Supplementary Fig. S5-S7). Once trained, the machine learning model can analyze new, unseen PA signals and accurately classify them into one of the five species categories based on the features it has learned.

To illustrate the effectiveness of our machine learning approach in classifying PA signals, principal component analysis (PCA) was utilized to visualize the testing data with both raw features and the features extracted by the machine learning model. PCA is a well-established technique for visualizing datasets by reducing their dimensionality. However, with only the raw features, the data of different species overlap considerably, making it challenging to differentiate the different species of red blood cells (Fig. 3h). In contrast, with the features extracted by our machine learning model, PCA reveals distinct decision boundaries between different categories, clearly separating the different red blood cell species, as shown in Fig. 3i. This demonstrates the significant improvement achieved by machine learning in classifying different types of blood cells, and the unique photoacoustic signatures can provide sufficient information to enable reliable classification.

These results demonstrate that our approach can not only detect the presence of free-flowing micro/nano particles in their natural environment but also obtain their photoacoustic properties. The frequency domain features of photoacoustic signals extracted by machine learning can be used to detect and classify different species of cells without the need for incubation, culturing, labeling, and imaging. We also characterized the PA signals on nanoparticle samples at varying concentrations. The PA signal intensity increased with higher nanoparticle concentrations, which can be used for quantitative detection (Supplementary Fig. S4). With these findings, we propose a vision for rapid, PA detection and characterization that allows for particle/cell identification without the need for full spectroscopic and/or microscopic analysis, which typically relies on bulky equipment and complicated sample preparation.

**Detection of cells in whole blood**

The design of mode protection within solid walls of the MBR ensures that the optical resonances for photoacoustic detection operate effectively without being affected by the solution's complexity. To explore the feasibility of sensing in complex matrix solutions, we tested whole blood samples of five different species—pig, sheep, turkey, goat, and horse, by injecting the sample into the sensor and recording the corresponding PA signals (Supplementary Video 1). Whole blood is a complex biological fluid with colloidal properties, comprised of multiple components including water, cellular elements (red blood cells, white blood cells, and platelets), various metabolites, dissolved electrolytes, a wide variety of proteins, and circulating hormones. The direct measurement of specific analytes in whole blood poses significant challenges for conventional sensor technologies. The presence of numerous biomolecules and cellular components can hinder the selective detection of target substances, leading to compromised accuracy and reliability in sensor performance. The blood samples tested in our experiments were diluted to 1% by volume, without any purification, labeling, or incubation process. Fig. 4a-e shows the time-domain and frequency-domain PA signals of whole blood from five different species. The PA signal variations among them are notably distinct, which can be used as fingerprints for identification. The spectral features in the PA signal are related to the optical absorption, density, thermal, and mechanical properties in the whole blood samples. Due to the complex composition of the blood matrix, relatively large deviations appear in the PA signal across measurements (Supplementary Fig. S8b and S9). To classify the different whole blood samples, robust algorithms are required to identify common features within the same kind of whole blood, while ensuring that these features are distinct enough to differentiate between different types of whole blood. Here, we implemented prototype learning, which maximizes the similarity of representations among samples from the same animal to extract decisive features for classification. Fig. 4f demonstrates the effectiveness of using PA signals obtained from whole blood samples for classification purposes. The clear and distinct decision boundaries between different categories highlight the ability to accurately differentiate between various sample types. Note that blood is a complex biological fluid containing various components that can interfere with conventional sensing techniques. However, the clear decision boundaries obtained from the PA signals suggest that this method is resilient to the inherent complexity of whole blood samples and allows us to accurately classify whole blood samples despite the inherent variability in the signal, making it a promising tool for blood-based analysis.

**Discussion**

In summary, we have developed an optofluidic sensing platform that integrates highly sensitive optical modes and the photoacoustic effect to achieve rapid, label-free, and high throughput measurements of particles. Unlike evanescent-field-based and SPR-based sensing platforms relying on surface binding for detection, our immobilization-free approach can detect free-flowing particles in fluid (away from the sensing surface), with excellent specificity achieved using PA spectroscopic signatures captured by a high-$Q$ MBR. By confining optical modes within the wall of the MBR sensor, this technique enables long-range acoustic-mediated measurements without reliance on random diffusion and spatially decouples the optical mode from the sensing volume, providing great resistance to potential contaminations in the solution. The measurement does not require external fluidic channels and chambers, lengthy culturing, expensive reagents, or thermal cycling equipment, and is robust to refractive index changes and absorption in the solution. This label-free and immobilization-free approach helps mitigate surface fouling, mass transport limitation, and surface regeneration challenges in conventional sensors. Furthermore, abundant information about the photoacoustic properties of particles can be obtained and implemented to classify the different morphologies of nanoparticles or different species of cells. Finally, our machine learning models achieve accurate classification, effectively distinguishing different types of particles and cells.

PA process exhibits intrinsic sensitivity to both the functional and molecular composition of the sample, leveraging the extensive optical absorption contrast present in biological systems. By choosing appropriate wavelengths to excite PA signals, we could selectively collect signals from specific objects. This unique capability allows for the non-invasive monitoring of a substantial number of red blood cells in their native physiological state when a 532 nm laser is used. Our approach circumvents the inherent complexities of whole blood, which often hinder conventional analytical techniques due to the potential for interference from the myriad of cellular and molecular components present in the blood matrix. Moreover, this technique minimizes the influence of external perturbation, such as temperature fluctuation and refractive index variations, to ensure an accurate assessment of the sensing targets through their PA signature.

In principle, this technique can be applied to the detection and characterization of a wide range of particles and cells in their natural states. Both artificial particles in nanomaterials or cells in biology can be measured with high throughput by choosing the proper wavelength of laser pulses or using a frequency comb[40] in the PA process[41–43]. For instance, features in the PA spectra can be further used to study the status of red blood cells and differentiate diseased cells from healthy ones. This would enable rapid disease diagnosis, hemoglobin C disease, for example, or hemoglobin S-C disease, sickle cell anemia, and various types of thalassemia[44,45]. Highly automated measurement and data acquisition processes hold promise for clinical and industrial applications. This technique offers versatility beyond liquid phase sensing, as it can also be applied to detect particles and molecules in the gas phase[46], providing great promise to a powerful sensing platform for various applications. When combined with large datasets, the AI-assisted sensing platform presented here could rapidly scan and identify cells in a patient sample and recommend an initial diagnosis in one step. It can also provide valuable particle information for nanomaterial characterization or environmental assessment without needing to wait for a culture or purification and random diffusion step. Such a smart sensing platform holds significant potential for advancing diagnostics, nanomaterial industries, and environmental monitoring.

## Materials and methods

### Sensor fabrication

MBRs were fabricated in line with a silica capillary (75 μm inner diameter and 125 μm outer diameter). First, we stripped away the polymer coating on the silica capillary using a butane torch and cleaned the silica window with isopropyl alcohol. Next, one end of the capillary was sealed with epoxy to allow the build-up of internal air pressure inside the capillary. The capillary was then placed onto a Vytran precision glass processing station and internally pressurized with air. The bare silica was locally heated and inflated into a spherical geometry – resulting in a microbubble resonator. The built-in microscope on the Vytran machine is used to monitor the fabrication process for quality control. By controlling the heating power and the applied pressure, the diameter as well as the wall-thickness of MBRs can be controlled.

### Experimental setup

Fig. 1b shows the experimental schematic for our MBR photoacoustic detection system. The built-in fluidic channels of the MBR are used to deliver the sample solution to the WGM resonator using a syringe pump. The pump was set to withdraw the solution through the MBR at a rate of 50 μL / min.

WGMs were excited in the microresonator using a tapered 780HP optical fiber from Thorlabs. We used an external cavity diode laser centered around 780 nm connected to a function generator to scan the wavelength of the laser around 40 pm at a rate of 60 Hz. An optical attenuator and polarization controller were used to adjust the light intensity and polarization before coupling into the microresonator to optimize the WGM spectrum. The transmitted intensity from the microresonator was detected by a high-speed photodetector connected to an oscilloscope.  Then we fixed the wavelength at an optical resonance by eliminating the scanning signal.

Photoacoustic excitation was achieved using a $Q$-switched 532 nm pulsed laser at a reputation rate of 60 Hz. This laser pathway is completely free space and focused on the MBR outer surface with an objective lens. An objective lens coupled to a CCD camera was used for optical alignment. This objective lens can also be scanned along the capillary axis, which means we are not limited to target detection just inside the MBR, but inside the capillary as well.

### Sample preparation

Four different geometries of AuNPs were used in this study, namely nanospheres, nanorods, nanocubes, and nanoshells. The AuNPs ranged in size from 10 nm to 50 nm and were suspended in a deionized water buffer with a citrate capping agent. The concentration of the colloidal AuNP ranged from $10^{10}$ nanoparticles / mL (spheres, rods, and cubes) to $10^{12}$ nanoparticles / mL (shells).

The five different species of whole blood as well as washed red blood cells were purchased from a commercial supplier (LAMPIRE Biological Laboratories). The whole blood was diluted to 1% in PBS solution and then delivered into the core of the sensor using a small syringe with a syringe pump.  The suspensions of washed red blood cells were also diluted to 1% in PBS solution.

### Scanning transmission electron microscopy and imaging

For STEM images, 6 µL of different AuNP solutions were deposited on a pure carbon film having a mesh size of 400 and were then dried for 60 minutes at ambient temperature (~21 °C). The carbon surface adsorbed the AuNPs particles and the water was evaporated. This carbon support film for STEM is very thin (15-25 nm) and highly transparent to electrons. The dried film is then mounted on the STEM sample holder for imaging.

**Machine learning of the PA signals**

Details of the dataset and machine learning analysis can be found in Supplementary Session 5 (Fig. S5). For AI analysis for feature learning and particle classification, we employed a 1-dimensional Convolutional Neural Network as a feature extractor to learn useful information about the photoacoustic signal in the frequency domain. In the convolutional layer, multiple convolutional kernels stride along the vectors of the input, where each kernel can capture a unique local pattern in the spectrum, and subsequent pooling layers distill essential features. Then the feature maps, after the pooling process, are flattened into a vector and propagated into a fully connected layer with the activation function of Rectified Linear Unit (ReLU). Later, this learned feature was used for making predictions based on prototype learning. The detailed results of the machine learning models can be found in Supplementary Fig. S6-S7.

**Prototype learning for robust prediction**

A potential limitation of CNNs is that they tend to learn surface statistical regularities in the dataset rather than higher-level abstract concepts[47]. For highly sensitive optical sensors like WGM, trivial environment changes can be detected, and a well-trained CNN may misclassify with slight perturbations of the spectrum. Therefore, a robust machine learning model is required to assist the detection and sensing.

In prototype learning[39], prototype embeddings represent different classes in the latent space. The classification of a sample is simply implemented by finding the nearest prototype embedding using Euclidean distance in the latent space. The prototype embeddings are denoted as $m_{yj}$ where $y \in \{1, 2, \ldots, C\}$ represents the index of the classes and $j \in \{1, 2, \ldots, K\}$ represents the index of the prototype embeddings in each class. In our work, the number of prototype embeddings of each category is set to be $K = 1$, assuming that for each class there is only one representative embedding $m_y$ in the latent space. These prototype embeddings, with a dimensionality equivalent to the feature space, can be initialized with trainable parameters so that they can be simultaneously updated with the model parameters in the training process.

For feedforward propagation, the prediction is determined by the distance between the sample $x$ and the prototype embedding $m_y$ in the latent space instead of calculating the $Softmax$ function:
$$x \in class \min_i ||f(x,q) - m_y||$$
where $f(x,q)$ denoted the output of the machine learning model (learned features) with trainable parameter $q$. Then for the loss function, the probability of the prediction result is proportional to the negative distance $- ||f(x,q) - m_y||$. Considering the non-negative of the probability and sum-to-one properties, this prediction can be written as:
$$p(y|x) = \frac{e^{- ||f(x,q) - m_y||}}{\sum_k e^{- ||f(x,q) - m_k||}}$$
Therefore, a cross-entropy (CE) loss based on prototype learning can be described as:

$$CE = -\log p(y|x)$$

Moreover, the robustness can be interpreted as that the learned feature is close to the prototype embedding in the latent space, which indicates that the model can neglect the noise and only preserve the key features of the class. Therefore, the loss function of Prototype Loss (PL) can be defined as minimizing the distance between the learned feature and the prototype embedding with the correct category:

$$PL = ||f(x, q) - m_y||^2$$

Eventually, the total loss can be defined as the sum of Cross Entropy Loss (CE) and the Prototype Loss (PL):

$$loss = CE + \lambda \times PL$$

where $\lambda$ is the hyperparameter that governs the influence of Prototype loss on feature extraction and decision boundary formation. As $\lambda$ gets larger, the embedding of extracted features will get closer to the prototype. In the experiment, the hyperparameter $\lambda$ is set to be 0.1.

## Acknowledgments

This project is supported in part by the Chan Zuckerberg Initiative (CZI). The authors acknowledge the Institute of Materials Science and Engineering (IMSE) at Washington University in St. Louis for the use of instruments, financial support, and staff assistance.

## Conflict of interest

L.Y. is a co-founder of DeepSight Technology, Inc., and has an equity interest in the company. She also serves as the chief technology officer of the company. The other authors declare no competing interests.

## Contributions

J.L. and M.A. contributed equally to this work. J.L., M.A., and L.Y. conceived the idea and designed the experiments. J.L. and M.A. performed the experiments to characterize the sensor performance. H.Y. developed machine learning models to classify nanoparticles under the guidance of C.L. D.K.S. collected nanoparticle images. All authors contributed to analyzing the results and writing the manuscript. L.Y. and C.L. supervised the project.

## Data availability

The data sets generated and/or analyzed during this study are not publicly available due to privacy and licensing restrictions. However, they can be made available from the corresponding author upon reasonable request.

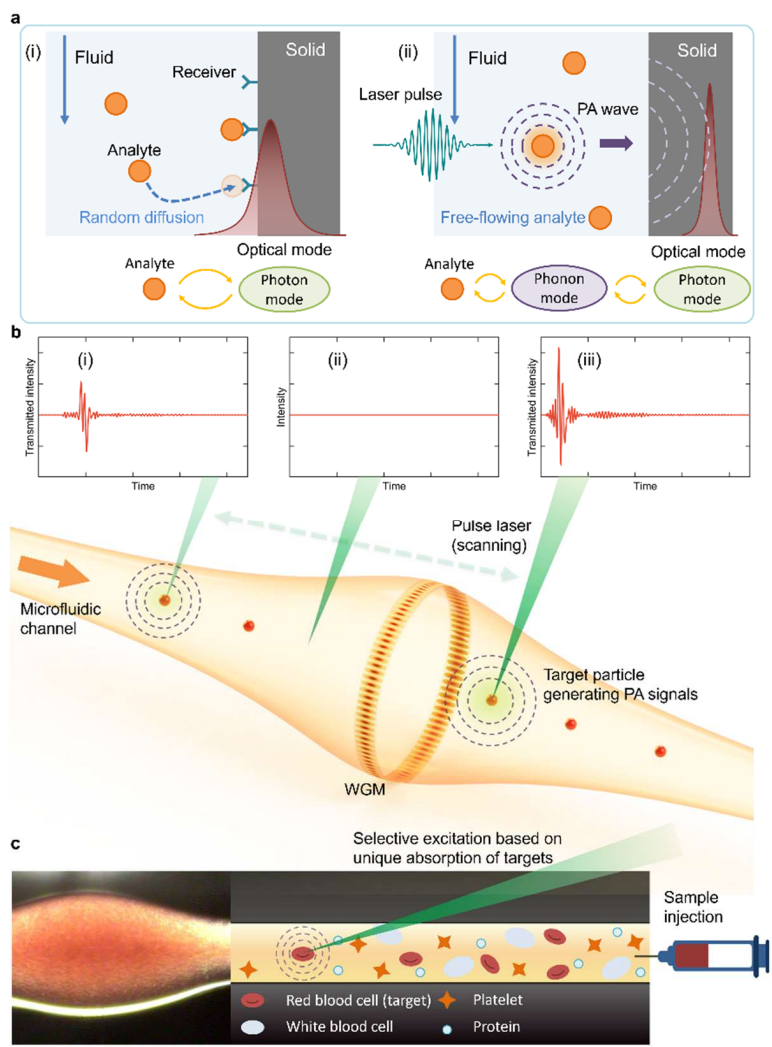

**Figure 1: Label-free all-optical photoacoustic (PA) microresonator-based sensors. a,** Principle of long-range acoustic-assisted sensing. (i) Conventionally, particle detection through optical sensors is achieved by directly perturbing the photon mode (or hybrid mode by coupling with a plasmonic mode). These evanescent-field-based methods require the binding of particles on the sensor. Analytes flowing in the fluid cannot be detected effectively. (ii) In our approach, upon exposure to a pulse laser, analytes with matching absorption generate photoacoustic waves propagating in the fluid medium. These PA waves can be efficiently captured by the high-$Q$ optical mode confined within the dielectric structure. The acoustic phonons mediate a long-range interaction between light and the analyte particles flowing in the microfluidic channel. **b,** Schematic of the sensing platform using PA in an optofluidic microresonator. The red particles indicate analytes flowing inside the microfluidic channel. The green pulse laser excites the analytes at different locations, which can be captured by the whispering gallery mode (WGM) confined in the equator of the resonator. (i) PA signal from an analyte far away from WGMs. (ii) No PA signal is detected in the absence of the analyte. (iii) A stronger PA signal from an analyte close to WGMs. **c,** Selective detection of red blood cells in a complex blood matrix. The left figure is a microscopic image when the microsensor is filled with whole blood. The right figure illustrates the specific excitation of PA signals in red blood cells in the presence of other cellular elements and proteins due to the unique absorption of hemoglobin in red blood cells at 532 nm.

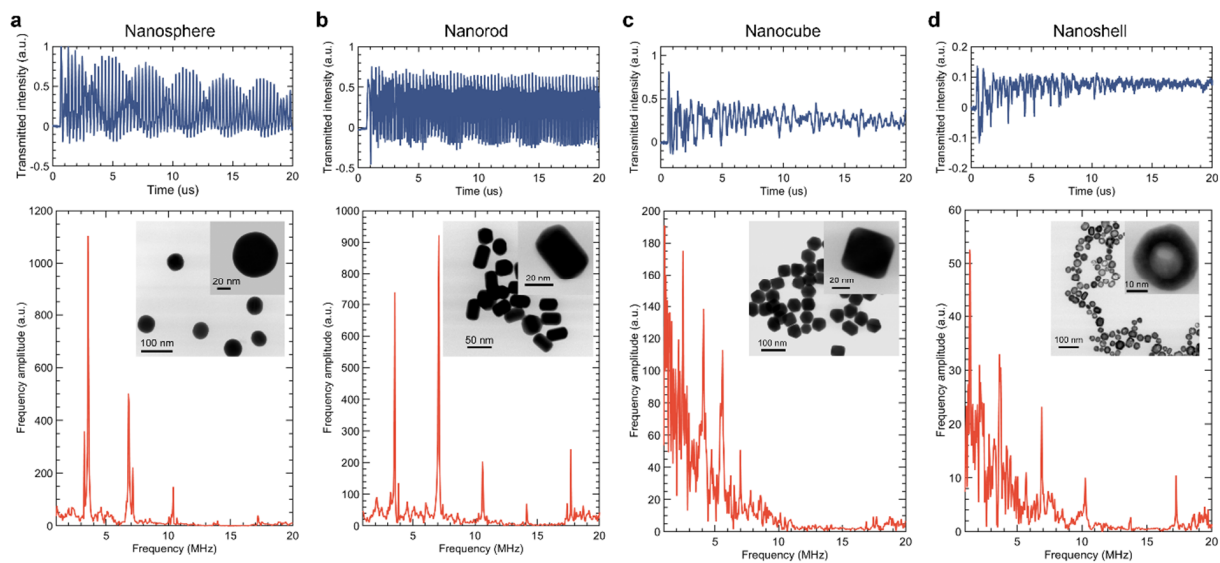

**Figure 2: Temporal and spectral measurement of gold nanoparticles (AuNPs). a-d,** Time domain photoacoustic signals and corresponding frequency domain spectra of AuNP from four different geometries. In each panel, corresponding scanning transmission electron microscope images are shown with zoomed-in insets.

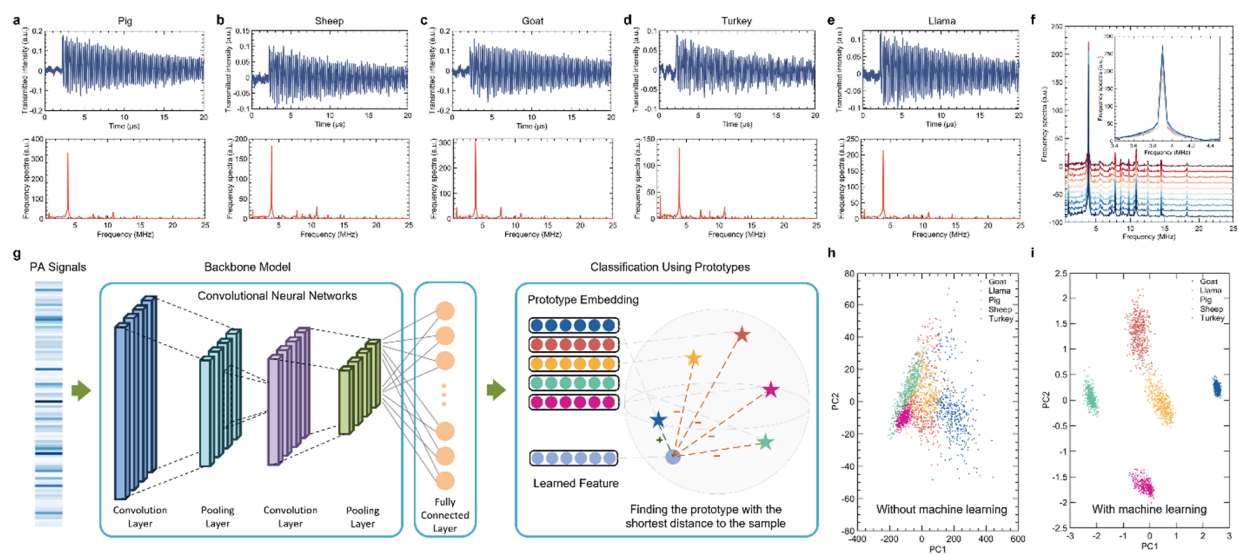

**Figure 3: Photoacoustic signals of red blood cells and machine learning analysis. a-e,** Time domain photoacoustic signals with calculated frequency domain of red blood cells from five different species of animals: pig, sheep, goat, turkey, and llama. The fast Fourier transform (FFT) spectra show very little difference between different kinds of cells. **f,** Offset FFT spectra of 10 individual measurements of red blood cells from the same species (sheep). The vibration frequencies in the PA signal remain consistent across all measurements. Inset: Close-up of the peak near 3.9 MHz across different measurements without offset. **g,** Schematic overview of the classification process and convolutional neural network (CNN) with prototype learning structure. FFT is performed on PA signals, followed by processing through the CNN. The CNN learns specific signal features during training, and then these features are used for classification by finding the nearest prototype embedding in the latent space. The model is eventually evaluated on the test set. Principal Component Analysis (PCA) is employed for feature visualization by simplifying the data into lower dimensions. **h,** PCA visualization of samples without machine learning-based feature extraction. The resulting data points are intermingled, making them difficult to distinguish. **i,** PCA visualization of samples using features extracted by the machine learning algorithm. The data points from each species form distinct, well-separated clusters. The results indicate that the machine learning-extracted features are effective in differentiating between the blood cells of various species compared to using raw signal features.

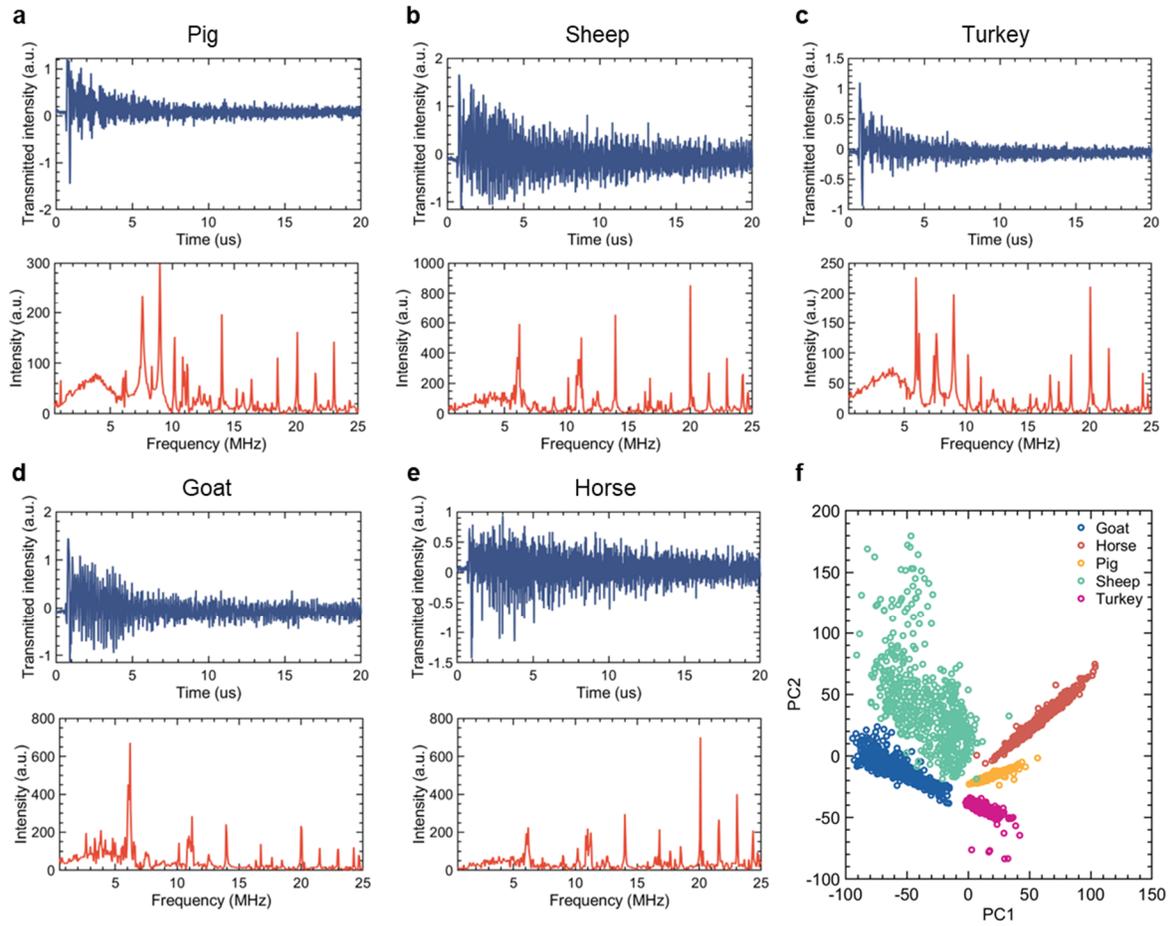

**Figure 4: Photoacoustic fingerprinting of whole blood samples. a-e,** Time domain photoacoustic signals and corresponding frequency spectra of whole blood samples from five different species of animals: pig, sheep turkey, goat, and horse. **f,** PCA visualization of samples using features extracted by the machine learning algorithm, indicating clear decision boundaries and clustering of each species in whole blood samples.

# Supplementary Information in Whispering-Gallery-Mode Resonators for Detection and Classification of Free-Flowing Nanoparticles and Cells through Photoacoustic Signatures

Jie Liao[1], Maxwell Adolphson[1], Hangyue Li[2], Dipayon Kumar Sikder[1], Chenyang Lu[1,2,3,4], Lan Yang[1,4]

[1]Department of Electrical and Systems Engineering, Washington University, St. Louis, MO 63130, USA
[2]Department of Computer Science and Engineering, Washington University, St. Louis, MO 63130, USA
[3]Department of Medicine, Washington University, St. Louis, MO 63130, USA
[4]AI for Health Institute, Washington University, St. Louis, MO 63130, USA
E-mail: yang@seas.wustl.edu

## 1. Sensing mechanism of long-range acoustic-mediated sensing

Conventional resonance shift sensing is a straightforward method to measure the target of interest by tracking induced changes in the resonance wavelength (or frequency) of an optical resonator. Fig. S1 (a) illustrates the resonance shift induced by the capture of target molecules on the resonator surface in many biosensing applications. This shift measurement allows for the extraction of quantitative and kinetic information about the binding of molecules. However, this method relies on random diffusion processes to bring particles to the sensing surface within the evanescent field, resulting in limited detection efficiency and randomized particle arrival locations. Consequently, only statistical or binary measurements can be obtained, and computation-intensive techniques must be employed for more detailed information. Furthermore, the optical properties of the modes are directly modified by the analyte, making this method unsuitable for detecting large amounts of particles that exhibit scattering or absorption. Additionally, environmental factors can also induce resonance shifts, potentially affecting the sensing accuracy.

In contrast, acoustic-mediated sensing operates at a fixed wavelength and measures the transmission intensity changes induced by acoustic waves generated by the analyte, as shown in Fig. S1 (b). The acoustic waves propagate through the sample fluid and are detected by the optical mode, eliminating the reliance on random diffusion processes. The intrinsic photoacoustic and mechanical properties of the analyte are collected in the acoustic waves. In this method, the detection occurs through photon-phonon interaction, and the optical mode remains confined within the resonator without directly interacting with the analyte. As a result, acoustic-mediated sensing is not limited by the optical scattering and absorption of analytes or the fluidic medium. Moreover, since the photoacoustic process produces ultrasound in the MHz frequency range, the detection is background-free, providing a high signal-to-noise ratio and improved accuracy.

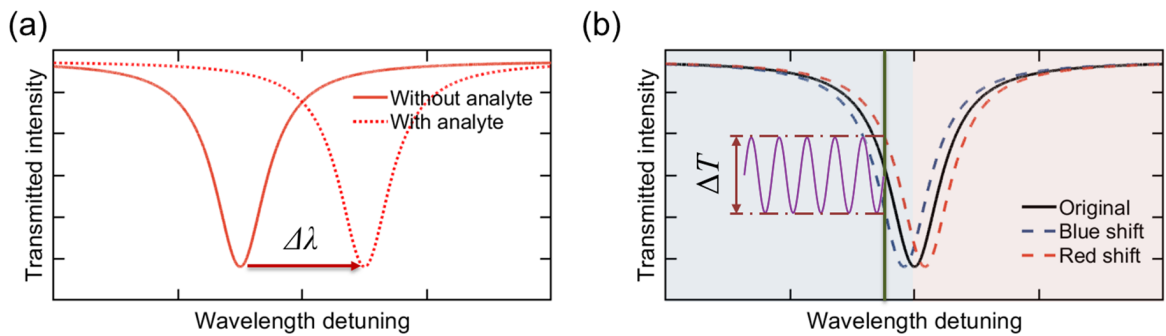

**Figure S1: Comparison of sensing mechanisms in optical microresonators.** (a) Conventional resonance shift sensing mechanism. The effective refractive index of the optical mode is modified by the binding of the analyte, resulting in the change of resonance wavelength. (b) Acoustic-mediated sensing. The acoustic wave modulates resonance and induces optical transmission modulation $\Delta T$ at a fixed wavelength.

## 2. Excellent stability for reliable measurement

Since the light field is confined within the thick wall of the microbubble resonator (MBR), direct interaction between whispering gallery modes (WGMs) and the sample solution is prevented. Consequently, when the core filling switches from deionized (DI) water to black dye, there's negligible change observed in the WGM spectra. This means that the $Q$-factor, along with the Signal-to-Noise Ratio (SNR), and therefore the overall sensing performance, remain unaffected by the composition and refractive index of the sample media itself. This characteristic is critical for the sustained effectiveness of the sensors in various sensing applications over an extended period.

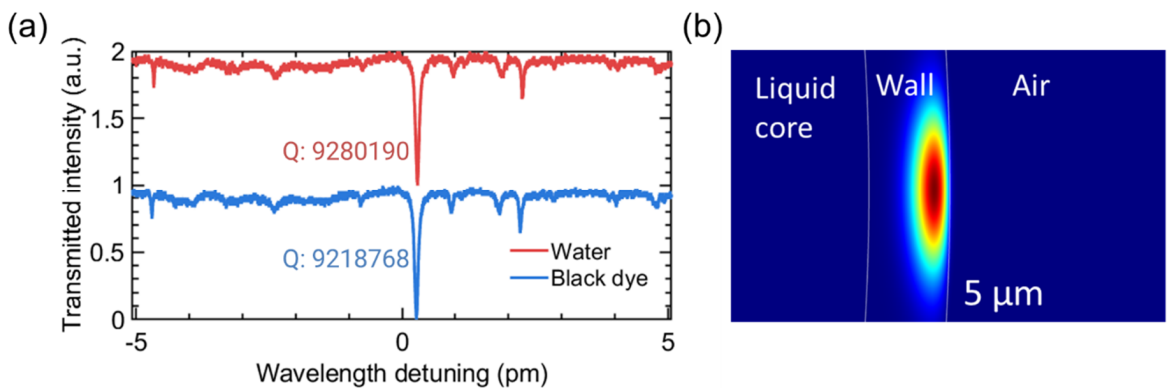

**Figure S2: Thick wall for mode protection.** (a) WGM spectra when the core of the sensor is filled with DI water and black dye, respectively. (b) COMSOL simulation of field distribution of a WGM. When the wall is 5 μm thick, there is little overlap between the WGM and the liquid core in the sensor.

## 3. Extended sensing range

By adjusting the beam spot position of the pulsed laser, we achieve an extended sensing range, enabling the detection of particles even at distances far from the microresonator. Fig. S3 (a) illustrates the scanning of the pulse laser along the transparent capillary that serves as the microfluidic channel. To experimentally validate this expanded sensing range, we moved the pulsed laser's beam spot along the microfluidic channel while keeping the laser current constant. The distance between the WGM and the particles causes a delay between the pulse excitation and the optical readout. The measured delay time changes linearly with the distance of the pulse laser excitation location from the WGM sensor, as shown in Fig. S3 (b). This delay time can be used as an indicator providing us with the position information of the particles. The signal amplitude, characterized as the peak-to-peak amplitude, drops at a larger distance. Even with the presence of loss, we can still obtain measurable signals at a very large distance of 6 mm. This capability, despite the inherent loss typically associated with the mechanical property of sample media and the capillary, underscores the extended reach of our sensing approach, making it particularly effective in scenarios where proximity to the target is a challenge.

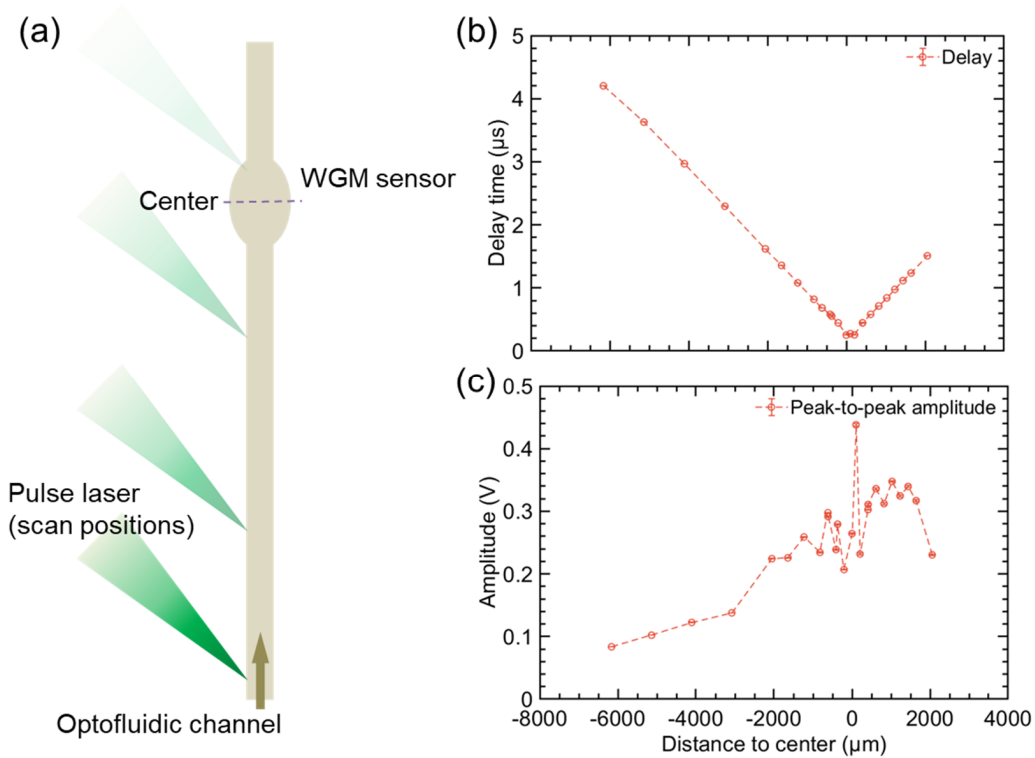

**Figure S3: Extended sensing range.** (a) Illustration of the scanning of the pulsed laser along the microfluidic channel, with WGMs being excited near the MBR equator while the sample solution flows through its core. (b) Time delays observed between the pulse trigger and the detected peak of the PA signals as the laser beam spot traverses along the channel. (c) Corresponding peak-to-peak amplitudes of the measured signals during the scanning.

## 4. PA signal amplitude as a function of nanoparticle concentration and input laser power

The photoacoustic response observed from the gold nanoparticles varies with the concentration of the nanoparticles in the solution and the input power delivered by the pulsed laser. Fig. S4 (a) shows that the photoacoustic signal amplitude increases with increasing concentrations of gold nanospheres. The dilutions were made from the original concentration of 3.28e10 nanoparticles / mL (nps/mL) and the concentration of the diluted samples was confirmed with UV/Vis spectroscopy. The power of the pulsed laser was held constant at 37 μW and a repetition rate of 60 Hz. We see a near linear relationship between the nanosphere concentration and the PA signal amplitude until about 3e10 nps / mL where saturation of the detector response is observed. In Fig.

S4 (b) we show the dependence of the PA signal amplitude on a sample of gold nanospheres at a constant concentration of 3.28e10 nps/ mL under increasing power from the pulsed laser.

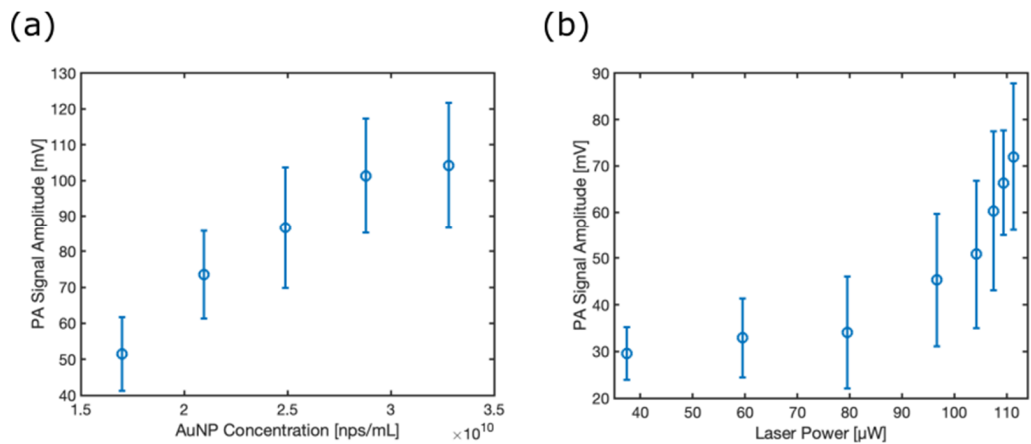

**Figure S4: PA signal amplitude as a function of nanoparticle concentration and input laser power.** (a) The PA signal response of gold nanospheres as the concentration is varied from 1.7e10 nps / mL to 3.28e10 nps / mL. (b) The PA signal response of gold nanospheres at a concentration of 3.28e10 nps / mL under increasing power delivered by the pulsed laser.

## 5. AI analysis for feature learning and particle classification

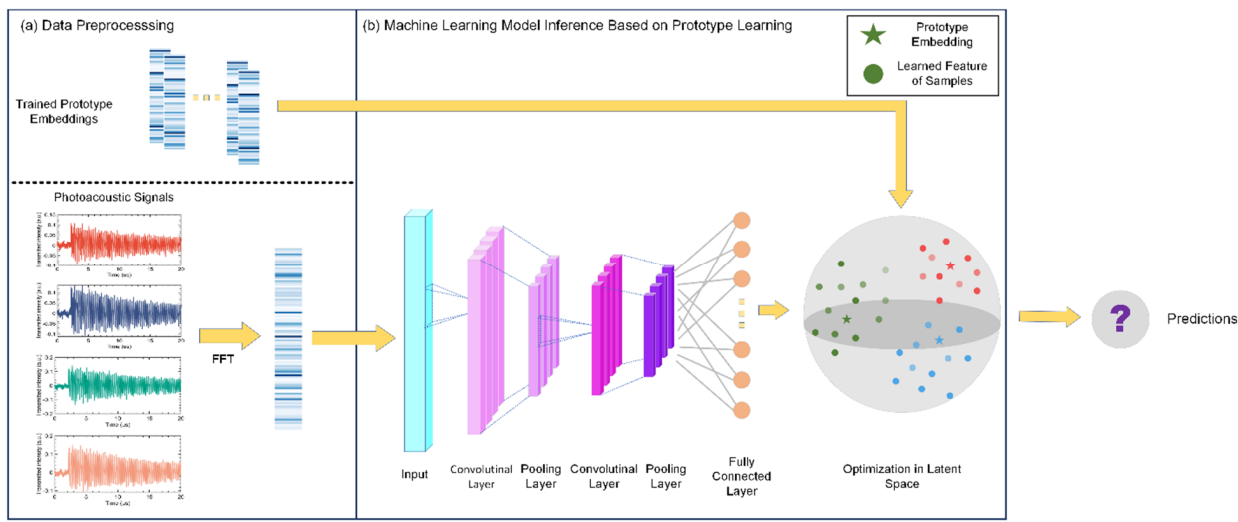

**Figure S5: Machine learning pipeline for classifying PA signals.** (a) The PA signals are preprocessed with Fourier Transform as input data. The learned prototype embeddings of the different classes are used for inference. (b) The backbone model comprises a 2-layer CNN and a fully connected layer. The distance between the learned features of the sample and the prototype embeddings of each class is calculated in the latent space. Finally, the sample is classified as the class of its closest prototype embeddings.

### 5.1 Data description and preprocessing

Two datasets are collected to train and test our machine learning model. The two distinct datasets allow us to demonstrate the generality of the machine learning approach.

**Reb blood cells:** The first dataset contains five different blood cells for animals: goat, llama, pig, sheep, and turkey. For each species, 1000 photoacoustic signals were collected such that the whole dataset contains 5000 different samples. For each sample, we use Fourier Transform to transform the PA signals to the frequency domain keeping the amplitude information as input features for the machine learning model. After preprocessing, the dimensionality of the input for machine learning is 10000. The dataset is separated into the training and testing sets with a ratio of 4:1, i.e., 8000 samples are used in training the model and 2000 samples are used to evaluate the model performance.

**AU nanoparticles:** The second dataset contains four different shapes of Au nanoparticles: nanocube, nanorod, nanosphere, nanoshell. For each category, 1000 photoacoustic signals are collected. We use a similar methodology as that used for the blood cells to process the AU nanoparticle data. We transform the PA signals to the frequency domain using Fourier Transform while also keeping the amplitude information. After preprocessing, the dimensionality of the input for machine learning is 10000. The dataset is separated into the training and testing sets with a ratio of 4:1, i.e., 8000 samples are used in training the model and 2000 samples are used to evaluate the model performance.

## 5.2 Model development

### Convolutional neural network for photoacoustic spectra

Convolutional Neural Networks (CNNs)[1] is widely adopted for analyzing spectroscopy[2–4]. CNNs are specifically designed to learn spatial hierarchies of features from input data. This feature extraction capability is particularly beneficial for spectra, which often contain complex signatures that are difficult to manually engineer and identify.

For feature extraction, we employ a 1-dimensional CNN to extract useful information from the photoacoustic signal in the frequency domain. In the convolutional layer, four convolutional kernels stride along the vectors of the input, where each kernel captures a unique local pattern in the spectrum. The subsequent pooling layers distill essential features and reduce the computational cost. Then, the feature maps generated after the pooling process are flattened into a new vector and propagated into a fully connected layer with the activation function of Rectified Linear Unit (ReLU). After the activation function, the obtained vectors are the learned features to be used for classification in prototype learning.

### Prototype learning for robust prediction

A potential limitation of CNNs is that they may learn surface statistical regularities in the dataset and cannot perform well with perturbations of input data such as noise[5]. A highly sensitive optical sensor (e.g., WGM) can detect small environmental changes. Consequently, small variations of the spectrum may lead to misclassifications by a well-trained CNN.

To enhance the robustness of the model, we implement prototype learning[6] (as described in Section Method). The prototype of the class is defined as the most representative embedding of the class in the latent space. In the training process, the model parameters and the prototype embeddings are simultaneously updated according to the gradient of the loss function.

With the learned embeddings of different classes, we classify new PA signals based on the distances from the learned feature of the PA signals to the prototypes of different classes in the latent space, as displayed in Fig. S5. The sample is classified as the class associated with the prototype that has the shortest distance to the sample's learned features.

**Details of hyperparameters and implementation**

The model is trained with 100 epochs and a learning rate of 0.001. Our training process also incorporates early stopping and a dropout rate of 0.3 to mitigate the effects of overfitting. The number of convolution layers is 2, and the stride of the convolution kernel and the stride of the pooling layer are set to 20. The number of the convolutional kernels is set to 4, i.e., the convolutional layer shares 4 different sets of parameters and can obtain 4 channels in the feature learning process. In the latent space, the dimensionality of the learned features, as well as the prototype embedding, is set to 64. For prototype learning, we train one prototype for each class. The model is programmed in Python 3.9.16 and Pytorch 1.8.0 and trained on RTX 3090 GPU, with a VRAM of 24 GB.

**5.3 Model performance and evaluations**

**Evaluations on various metrics:** To evaluate the performance of classifying the Au nanoparticles and the red blood cells, we made predictions on the testing data and measured these prediction outcomes with various metrics.

**Accuracy:** The proportion of correct predictions out of all predictions from the testing set.

**Macro recall:** The metric of recall is originally applied to binary classifications, which calculates the ratio of covered positive samples by the model: $\frac{True\_Positive}{True\_Positive+False\_negative}$. For classifications of multiple categories in our experiments, we use the macro recall defined as the mean value of recall in each class where the corresponding class is considered the positive label.

**Macro precision:** Similarly, the metric of precision is originally applied for the binary classification, which calculates the ratio of correctly predicted positive samples by the model: $\frac{True\_Positive}{True\_Positive+False\_Positive}$. For classifications of multiple categories in our experiments, we use the macro precision defined as the mean value of precision in each class where the corresponding class is considered the positive label.

**Macro F1 score:** The F1 score is the harmonic mean of precision and recall. The Macro F1 Score treats all classes equally, regardless of their frequency in the dataset.

The performance of each dataset is summarized in the following table:

**Table 1: Overall Performance of machine learning on two tasks**

| Metrics | Accuracy | Macro Recall | Macro Precision | Macro F1 |
|---|---|---|---|---|
| Au Nanoparticles | 0.9961 | 0.9956 | 0.9971 | 0.9963 |
| Red Blood Cells | 0.9870 | 0.9863 | 0.9867 | 0.9864 |

The machine learning models achieve excellent predictive performance in classifying the PA signals in both datasets. The details of the result for each category are shown in the following confusion matrix:

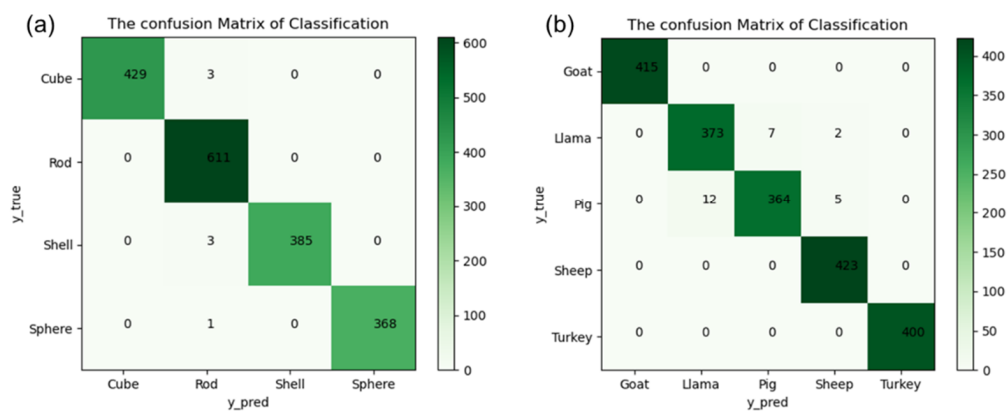

**Figure S6: Confusion matrices of the machine learning models.** (a) Result on Au nanoparticles. (b) Result on red blood cells.

## Ablation study on prototype learning

We introduced prototype learning to make robust classifications. To assess the effectiveness of prototype learning, we compare the performance with and without prototype learning. The confusion matrices of the two approaches are displayed in the following:

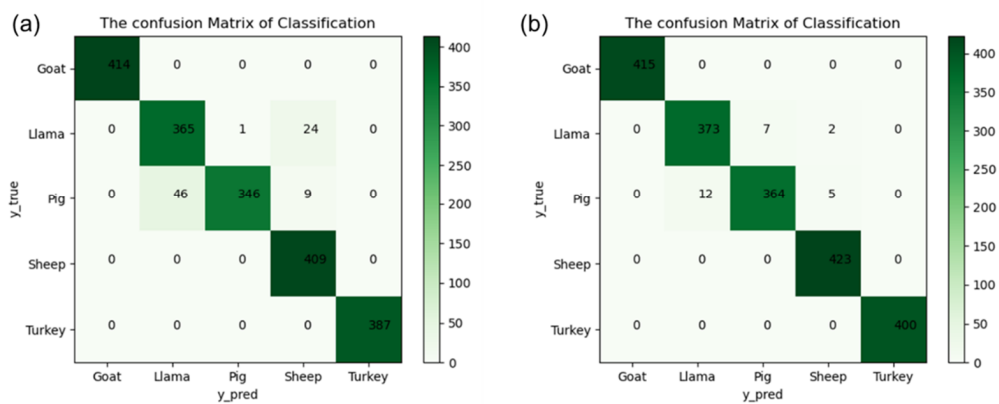

**Figure S7: Confusion matrix on red blood cells.** (a) Result without prototype learning. (b) Result with prototype learning.

The original CNN struggles to distinguish the red blood cells of pig and llama due to their significant similarities. In contrast, prototype learning enhances the model's ability to distinguish between llama and sheep.

## 6. Principal Component Analysis (PCA) without prototype learning

We performed PCA directly on the PA signals of red blood cells and whole blood samples, without the aid of AI. With only the raw features, the data of different species overlap considerably, making it challenging to differentiate the different species.

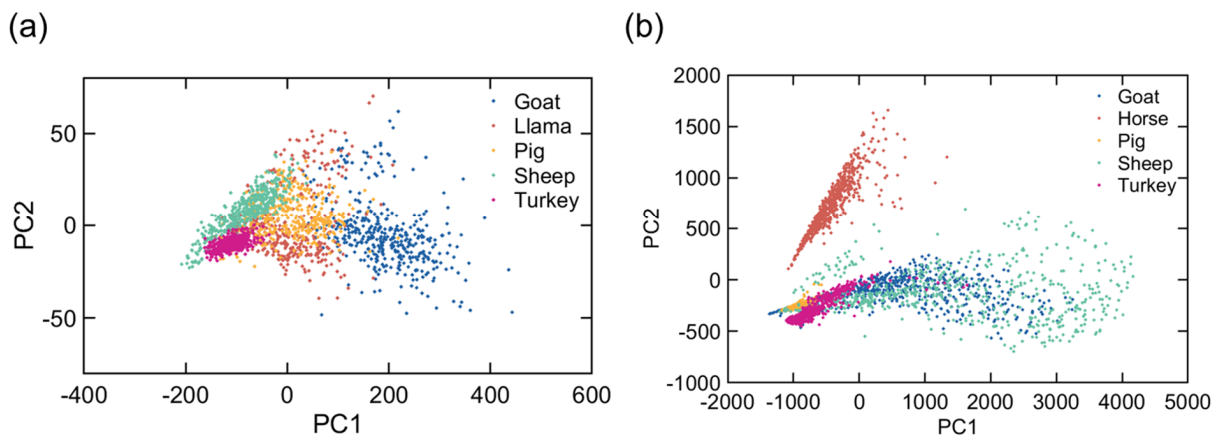

**Figure S8: PCA without prototype learning.** (a) PCA on input features from red blood cell signals. (b) PCA on input features from whole blood signals. Without prototype learning, the resulting data points are intermingled, making them difficult to distinguish.

## 7. Deviations in signals from whole blood samples

The PA signals obtained from whole blood samples exhibit larger deviations compared to those from red blood cell samples. These increased deviations can be attributed to the complex composition of whole blood. In addition to red blood cells, whole blood contains various other components such as white blood cells, platelets, and plasma, each of which may contribute to the PA signal in different ways. The presence of these additional components introduces more variability in the PA signals, leading to larger fluctuations and a wider range of signal intensities.

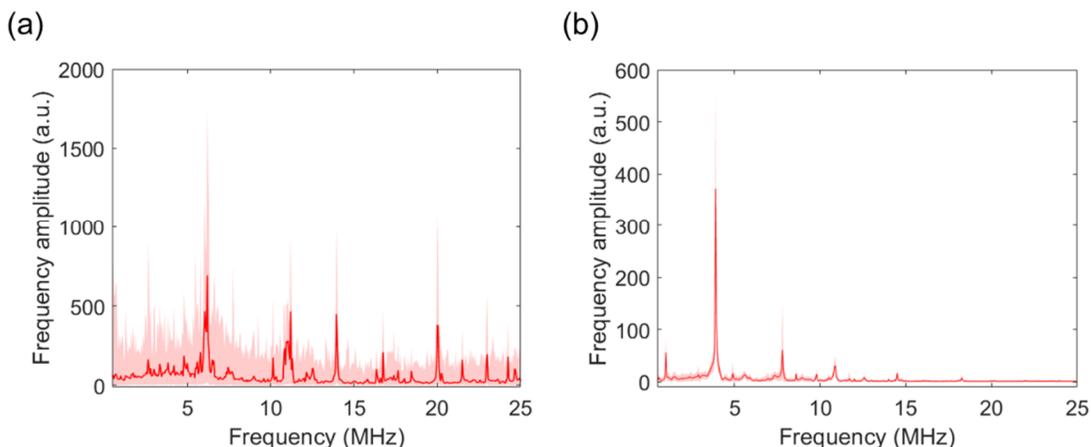

**Figure S9: Deviations in the PA signals.** PA signals from (a) whole blood and (b) red blood cell sample of goat. The red curve represents the average of 2000 frame signals. The pink-shaded area surrounding the red curve indicates the fluctuations in the PA signals, spanning from the minimum to the maximum values observed across the 2000 frames.